\documentstyle[12pt,graphicx]{article}
\title{{ Limits on SUSY Particle Spectra from Proton 
Stability and Dark Matter Constraints }}
\author{R. Arnowitt$^*$  
  and Pran Nath$^{**}$\\
$*$Center for Theoretical Physics, Department of Physics\\
Texas A \& M University, College Station, TX  77843-4242 \\
$**$Department of Physics, Northeastern University \\
Boston, MA 02115-5005}
\begin{document}
\maketitle  
\abstract{ It is shown that the combined constraints on the amount of 
cold dark matter and of proton stability produce a stringent upper 
limit on the gluino mass $m_{\tilde g}$ (and hence on the lightest
neutralino mass $m_{\chi_1^0}\simeq  m_{\tilde g}/7$ for a large 
class of gravity mediated  supergravity unified models. One finds that
for the minimal SU(5) model current data (Kamiokande) restricts 
$m_{\tilde g}\leq 400$ GeV for a scalar soft breaking mass $m_0\leq 1$ 
TeV. Expected future data from Super Kamionkande and ICARUS will be
sensitive to the entire range of gluino mass for $m_0\leq 1$ TeV, and
be able to exclude the region $m_{\tilde g}\geq 500$ GeV for $m_0\leq 5$
TeV. Effects of quark mass textures are studied and one finds that the
bound $m_{\tilde g}\leq 500$ GeV holds when the experimental proton 
lifetime for the $p\rightarrow \bar \nu K^+$ mode becomes 
$\geq 5\times 10^{32}$ yr. Implications of these results for a test of 
these models at the Tevatron and at the LHC are discussed. The effects
of non-universal soft breaking in the Higgs and the third generation 
squark sectors are also examined, and it is found that the proton lifetime 
is sensitive to these non-universal effects. The current data already 
eliminates some regions of non-universalities. The constraints of 
proton stability on the direct detection of dark matter are seen 
to reduce the maximum event rates by as much as a factor of $10^3$.}

\def\til#1{#1 \over {4 \pi}}
\def\Mz2{M_Z^2}
\def\mh{m_{1/2}}
\def\sms#1#2{m^2_{\tilde#1,#2}}
\newtheorem{theorem}{Theorem}

{\bf Introduction}: 
In this Letter we study the combined constraints of proton stability 
and supersymmetric dark matter on supergravity unified models  with
R-parity invariance. 
We consider models where gravity breaks supersymmetry in a hidden 
sector at scale $\geq M_G$ (the GUT scale) with gravity as the 
messenger  to the physical sector.  
Models with R parity invariance eliminate 
in a natural fashion dimension four operators responsible for rapid 
proton decay. However, as is well known  one can have 
proton decay proceeding via dimension five operators\cite{wein,acn}. 
Simultaneously,
with R parity invariance one finds that  the lightest 
supersymmetric particle (LSP) is absolutely stable and further in supergravity 
unification\cite{can,applied} the LSP is seen to be the 
lightest neutralino ($\chi_1^0$) in almost all of
the parameter space of the model\cite{scaling}, 
and thus the $\chi_1^0$ is a  candidate for cold dark matter (CDM).  
 In this    
Letter we show that the combined constraints of proton stability and
dark matter put an upper limit on the gluino mass of 500 GeV for the 
case of the minimal supersymmetric SU(5) model as well as 
constraining the masses of the other SUSY particles. We argue that this 
result would hold for a much wider class of models if the current lower
limit on the proton lifetime is increased by a factor of 4.
 We also 
 show that p stability is sensitive to the nature of 
 non-universal soft breaking\cite{soni,matallio,nonuni} 
 and that GUT models with certain  type of non-universal soft breaking
  can already 
 be eliminated on the basis of p stability and dark matter 
 constraints.
 
{\bf Proton Stability Constraint}:  
It is well known that  
 even with R parity invariance, instability of the 
proton can arise in most SUSY and
string models via dimension five operators due to 
the exchange of color triplet Higgsinos. In general the
Higgsino mediated p decay is governed by the interaction 
$\bar H_1 J+\bar K H_1+ \bar H_i M_{ij}H_j$ 
where J and K are matter currents defined in terms of quark and lepton
fields and we have carried out a field redefinition so that $H_1$ 
and  $\bar H_1$ 
are the Higgs combinations that couple to the currents. Then it is easily 
seen that the condition needed to suppress proton decay from dimension five
operators is given by
$ (M^{-1})_{11}$=0.
A constraint such as this can arise either as  a consequence of discrete
symmetries or as a consequence of non-standard choice of matter 
representations. However, most of the SUSY/string models do not exhibit a 
constraint of this type and  
consequently in such models
one will have proton decay induced by dimension five 
operators\cite{flipped}. One needs
a doublet-triplet splitting to achieve a heavy Higgs triplet 
to suppress dimension five proton decay and a number
of mechanisms have been discussed in the literature to accomplish this.
These include fine tuning, the sliding singlet mechanism valid for 
SU(N) with $N\geq 6$, the missing partner mechanism, VEV alignment, Higgs
as a pseudo-Goldstone particle and the choice of more than one adjoint to break  the
GUT symmetry. 
    For a wide class of models, i.e., SU(5), SO(10), E(6), etc where
   proton decay proceeds via dimension five operators, the dominant
   decay mode of the proton is given by $p\rightarrow \bar\nu K^+$ and 
   the decay width may be parametrized by\cite{acn} 
   \begin{equation}
   \Gamma (p\rightarrow \bar\nu K^+)=\sum_{i=e,\mu,\tau}\Gamma(p\rightarrow
   \bar\nu_iK^+)\nonumber\\
   =(\frac{\beta_p}{M_{\tilde H_3}})^2 |A|^2|B|^2C
   \end{equation}
   where $M_{\tilde H_3}$ is the Higgs triplet mass, $\beta_p$ is the 
   matrix element of the three quark operator between the vacuum and the
   the proton state and its  numerical value  from lattice gauge calculations
   is given by\cite{gavela} $\beta_p=5.6\times 10^{-3}$ GeV$^3$,
A depends on the quark masses and the  K-M matrix elements, B 
contains the dressing loop integrals which depend on the SUSY mass 
spectrum, and C contains  the chiral current algebra factors which 
convert the chiral quark Lagrangian into an effective Lagrangian 
involving mesons and baryons(Chadha et.al. ref.[1]). In the analysis
of this paper we constrain $M_{H_3}$ mass to satisfy the relation
 $M_{H_3}\leq 10 M_G$ as in ref.\cite{acn}. 

	We begin by analysing the proton lifetime within minimal 
	SU(5) supergravity
	with radiative breaking of the electro-weak symmetry. This
	model is
	characterized by the following parameters: the universal 
	scalar mass $m_0$, the universal gaugino mass $m_{1/2}$,
	the universal tri-linear coupling $A_0$, and $tan\beta$ which
	is the ratio of the two Higgs VEVs needed in SUSY models to break
	the electro-weak symmetry. (The extensions to non-minimal models
	will be discussed below). The 
	maximum lifetime of the proton is then calculated  
	as a function of the gluino mass within a given
	naturalness constraint on $m_0$.  The result
	is given in Fig. 1 where we plot the maximum $p\rightarrow \bar \nu K^+$
	lifetime for naturalness
	limits on $m_0$ of 1 TeV, 1.5 TeV, and 2 TeV. (We estimate an
	error perhaps of a factor of  2, 
	from uncertainties in quark masses, CKM 
	factors, $\beta_p$ etc).  One may compare
	these results with the current experimental limit on this
	decay mode of $>1\times 10^{32}$ yrs\cite{pdg} and with the lower limit 
	of 
$ \tau(p\rightarrow \bar \nu K^+)> 2\times 10^{33}$ yrs
expected from Super Kamiokande (Super K)\cite{totsuka} and 
a lower limit of $ \tau(p\rightarrow \bar \nu K^+)> 5\times 10^{33}$ yrs
 expected from ICARUS\cite{icarus}. 
One finds that
the expected sensitivity of the Super K will be able to 
exhaust essentially all of the parameter space of the minimal model 
except for gluino masses less than about 400 GeV for the case when
$m_0\leq 1$ TeV. Similarly for $m_0\leq 1.5$ TeV the expected lower
limit from Super K  will be able to exhaust the parameter
space of the model for $m_{\tilde g} < 750$ GeV.  For the case of
$m_0\leq 2$ TeV,  Super K  will not constrain the 
gluino mass within its naturalness limits of $m_{\tilde g}\leq 1$ TeV.
The constraints from ICARUS will be stronger if it can reach its maximum 
sensitivity of $5\times 10^{33}$ yr. Here one will be able to exhaust
the  gluino mass range $m_{\tilde g}\leq 1 $ TeV for $m_0\leq 1$ TeV,
the gluino mass range  $m_{\tilde g}\leq 450 $ GeV for   $m_0\leq 1.5 $ TeV, and the 
 the gluino mass range  $m_{\tilde g}\leq 800 $ for  $m_0\leq 2$ TeV.

{\bf Relic Density Constraints}:
As mentioned in the Introduction, SUSY unified models with R parity 
invariance imply the existence
of an LSP which is absolutely stable,
 and  that over most of the parameter space the LSP is the lightest 
  neutralino which is generally  a combination of the two neutral 
SU(2)$\times U(1)$ gauginos and the two neutral SU(2) higgsino states.
 There are various models  for dark matter. In 
our analysis the details of these  models do not play a significant
role. Rather we will show that for a broad class of models where
the neutralino is the CDM one finds rather stringent constraints on proton
stability analyses. The quantity of theoretical interest is 
$\Omega_{\chi_1^0} h^2$ where $\Omega_{\chi_1^0}=\rho_{\chi_1^0}/\rho_c$, 
and $\rho_{\chi_1^0}$ is the neutralino relic density, $\rho_c$ is
the critical relic density needed to close the universe, and h is the Hubble
parameter in units of 100 km/sMpc
The current astrophysical data allows one to impose the constraint 
 \begin{equation}
 0.1\leq \Omega_{\chi_1^0}\leq 0.4
\end{equation}
Theoretically one computes the neutralino relic density using the
Boltzman equation for the neutralino number density in the early 
universe, and  $\Omega_{\tilde\chi_1^0} h^2$ is obtained by integrating
 the number density from the current temperature to the 
freezeout temperature. One has\cite{jungman} 
\begin{equation}
\Omega_{\tilde\chi_1^0} h^2\cong 2.48\times 10^{-11}{\biggl (
{{T_{\tilde\chi_1^0}}\over {T_{\gamma}}}\biggr )^3} {\biggl ( {T_{\gamma}\over
2.73} \biggr)^3} {N_f^{1/2}\over J ( x_f )}
\end{equation}
~\\
Here  $x_f= kT_f/m_{\tilde{\chi}_{1}}$, where k is the Boltzman 
constant, $T_f$ is the freezeout  
 temperature, $ T_{\gamma}$ is the current micro-wave background 
 temperature,  $(T_{\tilde\chi_1^0}/T_{\gamma})^3$ is  the reheating
 factor,
   $N_f$ is the number of degrees of freedom at 
 freezeout,  and  $J~ (x_f)$ is given by
~\\
\begin{equation}
J~ (x_f) = \int^{x_f}_0 dx ~ \langle~ \sigma \upsilon~ \rangle ~ (x) GeV^{-2}
\end{equation}
~\\
\noindent
Here $<\sigma v>$ is the thermal average, $\sigma$ is  the
   annihilation cross-section for the neutralinos and $v$,  their
  relative velocity.  $\sigma$
  involves  cross channel sfermion exchanges and direct channel 
  exchanges of Z and Higgs bosons.  The annihilation of the neutralinos via 
  direct channel Z and Higgs poles will be seen to play an important role 
  in our analysis. It is thus important that one carries out the correct
  thermal averaging over these poles. In the analysis of this work we 
  have employed the accurate method for this purpose\cite{greist,accurate}.

We discuss next the implications of imposing simultaneously the proton 
stability and dark matter constraints. The result of the analysis for the 
case of the minimal SU(5) unification is
exhibited in Fig2. One finds that with the naturalness constraint of
$m_0=1$ TeV (solid curve)  and with the relic density constraint of  
   	 $0.1\leq \Omega_{\chi_1^0}h^2\leq 0.4$ 
the  upper limit on the gluino mass falls below
400 GeV when the current experimental limit on p lifetime of 
  $\tau (p\rightarrow \bar \nu K^+)>1\times 10^{32}$ is imposed. 
	In Fig.2 we have also  considered the case when 
   	 the naturalness limit on $m_0$ is raised  to 5 TeV (dashed curve). 
   	 We find that even in
   	 this case the upper limit on the gluino mass 
   	  does not exceed 500 GeV. 
     	This remarkable result arises because the 
   	relic density constraint of Eq.(4) requires that $m_0$ 
   	be small, i.e., $\leq 200 $ GeV, for gluino masses
   	$m_{\tilde g}> 450$ GeV. 
   	The reason for this is that in this region, the t-channel 
   	sfermion exchanges dominate the neutralino annihilation 
   	cross section, and large $m_0$ (i.e., large sfermion mass)
   	will not give sufficient annihilation to satisfy the upper 
   	bound of Eq.(2). However,  
   	since the proton decay rate has a rough 
   	dependence on $m_{\tilde g}$  and  $m_0$ of the form 
   	$m_{\tilde g}/{m_0^2}$\cite{acn}, large values of 
   	the gluino mass and small values of $m_0$ tend to destabilize
   	the proton. 
	  Thus unified models in the neutralino annihilation region 
	  beyond the  Z pole and the Higgs pole 
   	region, i.e., in the region $m_{\tilde g}\geq 450$ GeV, tend to
   	act like certain no-scale  models ($m_0=0$) which also exhibits
   	proton instability\cite{noscale}.  In Table 1 we give the factor
   	by which the proton lifetime in reduced due to the relic density
   	constraint of Eq.(4) as a function of the gluino mass. 
   	This reduction factor lies in the range 10-30 and is thus
   	very significant. We note that the reduction factor is largly
   	independent of the GUT physics which enters via the Higgs triplet
   	mass and also is independent of the naturalness assumption
   	on $m_0$ in this region of the gluino mass as already discussed above.  
   	An upper limit of 450 GeV for the minimal supergravity model 
   	may be testable at the upgraded Tevatron if it can reach its optimum
   	energy and luminosity, and  the minimal model will be fully 
   	tested at the LHC. 
       The simultaneous 
 	  constraints of p stability and relic density constraints 
 	  also effect the supersymmetric particle spectrum. An example 
 	  of this is shown in Fig.3 where the minimum and the maximum of
 	  the Higgs mass is plotted as a function of the gluino mass. 
 	  One finds that the Higgs mass has  a lower limit of 75 GeV 
 	  which is consistent with the current LEP data but a significant
 	  amount of the parameter space could be tested
 	  at LEP190. 
 	  
{\bf Non-minimal Models:}
   We discuss now the effects that
   non-universalities may  have on the results of our analysis.
   As is well known non-universalities in the soft 
   SUSY breaking sector introduce flavor changing neutral currents (FCNC) 
   which are strongly constrained by experiment. One kind of non-universality
   which is not very stringently constrained by FCNC is the non-universality
   in the Higgs sector. One may parametrize it at the GUT scale by 	
$m_{H_1}^2(M_G) =  m_0^2 (1 + \delta_1),  
~~m_{H_2}^2(M_G) = m_0^2 (1 + \delta_2)$ 
with a reasonable range of  
 $\delta_i$ given  by $|\delta_i|\leq 1$ (i=1,2).
However,
it was pointed out in ref.\cite{nonuni} that one should also consider
the non-universalities in the third generation sector along with the
non-universalities in the Higgs sector as the 
non-universalities in the Higgs sector and in the third generation sector 
give contributions of the same size and  may enhance or cancel each other. 
 We therefore parametrize 
non-universalities in the third generation sector by
$  m_{\tilde Q_L}^2=m_0^2(1+\delta_3)$, ~~$m_{\tilde U_R}^2=m_0^2(1+\delta_4)$
where as before one limits $|\delta_i|\leq 1$(i=3,4). 
The effects of non-universalities in the third generation sector are 
similar to those in the Higgs sector and we will not consider them in
detail here. Instead we will focus on the non-universalities in the Higgs
sector.
Fig.4 gives the result of the analysis for the case of 
$\delta_1=1=-\delta_2$. We find that the limit on the gluino mass 
consistent with the proton lifetime constraints
falls below   400 GeV and $\tau(p\rightarrow \bar \nu K)$ 
generally lies below the value of  the universal
case. We have carried out a similar analysis for the case of 
$\delta_1=-1=-\delta_2$. Here, however, we find that in the region
of $m_{\tilde g}<500$ GeV, radiative breaking generally  gives 
rise to a large
tan$\beta$ and a small $m_0$ resulting in a maximum proton lifetime 
already below the current Kamiokande proton lifetime upper limit. For values of
$m_{\tilde g}>500$ GeV the relic density constraint leads to 
theoretical upper limits on the p-lifetime which also fall below the 
current experimental limits for p-decay lifetime except in a couple
of narrow domains of the gluino mass. These narrow corridors will also 
be tested when the proton lifetime limit on the 
$p\rightarrow \bar\nu K$ mode improves by a factor of about
4 over  the current limit.

 	The analysis we presented above was for the SU(5) model.
 	Similar analyses should be valid for all 
 	unified models where one has low tan$\beta$, i.e., tan$\beta$
 	$\leq 25$ because the constraints of the relic density on $m_0$
 	 would not be substantially modified in this case. The SO(10) 
 	 models  typically have $b-t-\tau$
 	unification which implies large tan$\beta$, i.e., tan$\beta\sim$ 50.
 	A large tan$\beta$  tends to destabilize the proton 
 	requiring a large effective mass 
 	$M^{-1}_{11}$$\simeq 10^{17-18}$
 	GeV which tends to ruin the 
 	success of the unification of couplings\cite{barr,lucas,urano}
 	 unless one has large
 	threshold corrections in the GUT sector. For these reasons the
 	SO(10) case requires a separate treatment which is outside
 	the scope of this Letter.

	  Next we discuss the effects that inclusion of 
	  textures\cite{georgi,anderson,pn} will have 
 	  on the analysis. An appropriate  treatment of textures requires  
 	  inclusion of higher dimensional operators in the effective 
 	  potential such as those due to Planck scale corrections. A rough
 	  analysis shows that the $p\rightarrow \bar\nu K$ lifetime is 
 	  modified  by a factor $\sim (\frac{9}{8}\frac{m_s}{m_{\mu}})^2$.
 	  Effectively the p-decay lifetime is enhanced by a small numerical 
 	  factor ($\sim 3-5$). The relative suppression 
 	  due to the relic 
 	  density constraint remains essentially unchanged when the 
 	  enhancement factor is included.
 	  Including the relative suppression due to relic density 
 	  constraints as given in Table 1 one finds that the limit 
 	  on the gluino mass, $m_{\tilde g}\leq 500$ GeV  will occur 
 	  when the 
 	  experimental lower limit on the proton decay lifetime for
 	  the $p\rightarrow \bar\nu K$ mode approaches 
 	  $\sim 5\times 10^{32}$ yr. We  
 	  expect the results of our analysis also to hold for the Calabi-Yau 
 	  Models of the type  discussed in Ref.\cite{calabi}.

{\bf Implications for Dark Matter Detection}: 	  
 	  Finally we discuss implications of the constraints of proton 
 	  stability on the direct detection of neutralino dark matter,
 	  e.g., in the scattering of neutralinos from target nuclei. The
 	  result of the analysis is given in Fig.5 for the minimal 
 	  SU(5) model. One finds that the 
 	  effect of p stability constraint is to significantly reduce
 	  the maximum event rates curves, by as much as  a factor of 
 	  $10^3$. 
 	  The reason for this is because, as discussed  above, proton 
 	  stability generally requires a large $m_0$, which reduces 
 	  significantly the dark matter detection event rate.
 	  (The maximum neutralino mass of about 65 GeV for the dashed
 	  curve corresponds to the bound $m_{\tilde g}\leq 450 $ GeV in Fig 2.)
 	  We expect this typical suppression to hold also in 
 	  other unified models under the simultaneous 
 	  constraints of p stability and relic density
 	  constraints. The results imply that p stability constraint 
 	  renders the detection of dark matter significantly more difficult.
 	  Thus SUSY unified models which allow p decay via
 	  dimension five operators will require significantly more 
 	  sensitive dark matter detectors, more sensitive by a factor
 	  of $10^3$ or more than those currently available for the
 	  detection of supersymmetric dark matter\cite{cline}. \\

%\section*{Acknowledgements}
 This work was  supported in part by NSF grants 
 PHY-9722090 and PHY-9602074.

\begin{center} \begin{tabular}{|c|c|}
\multicolumn{2}{c}{Table~1: Reduction of $\tau(p\rightarrow \bar \nu K)_{max}$ 
from CDM constraint
   } \\
\hline
gluino mass (GeV)  & reduction factor when $0.1<\Omega h^2<0.4$  \\
\hline
  500 & 29.5   \\
\hline
 550 & 23.2  \\
\hline
 600 & 18.6  \\
\hline
 650 & 15.3   \\
\hline
 700  & 12.9   \\
\hline
 750 & 11.3   \\
\hline
 800 & 9.8  \\
%\hline
% 820 & 124  \\
\hline
\end{tabular} 
\end{center}

%%%%%%%%%%%%%%%%%%%%%%%%%%%%%%%%%%%%%%%%%%%%%%%%%%%%%%%%%%%%%%%%%%%%%  
\begin{figure}[htbp]
\begin{minipage}[t]{6.0in}
	\includegraphics[angle=0,width=6.0in]{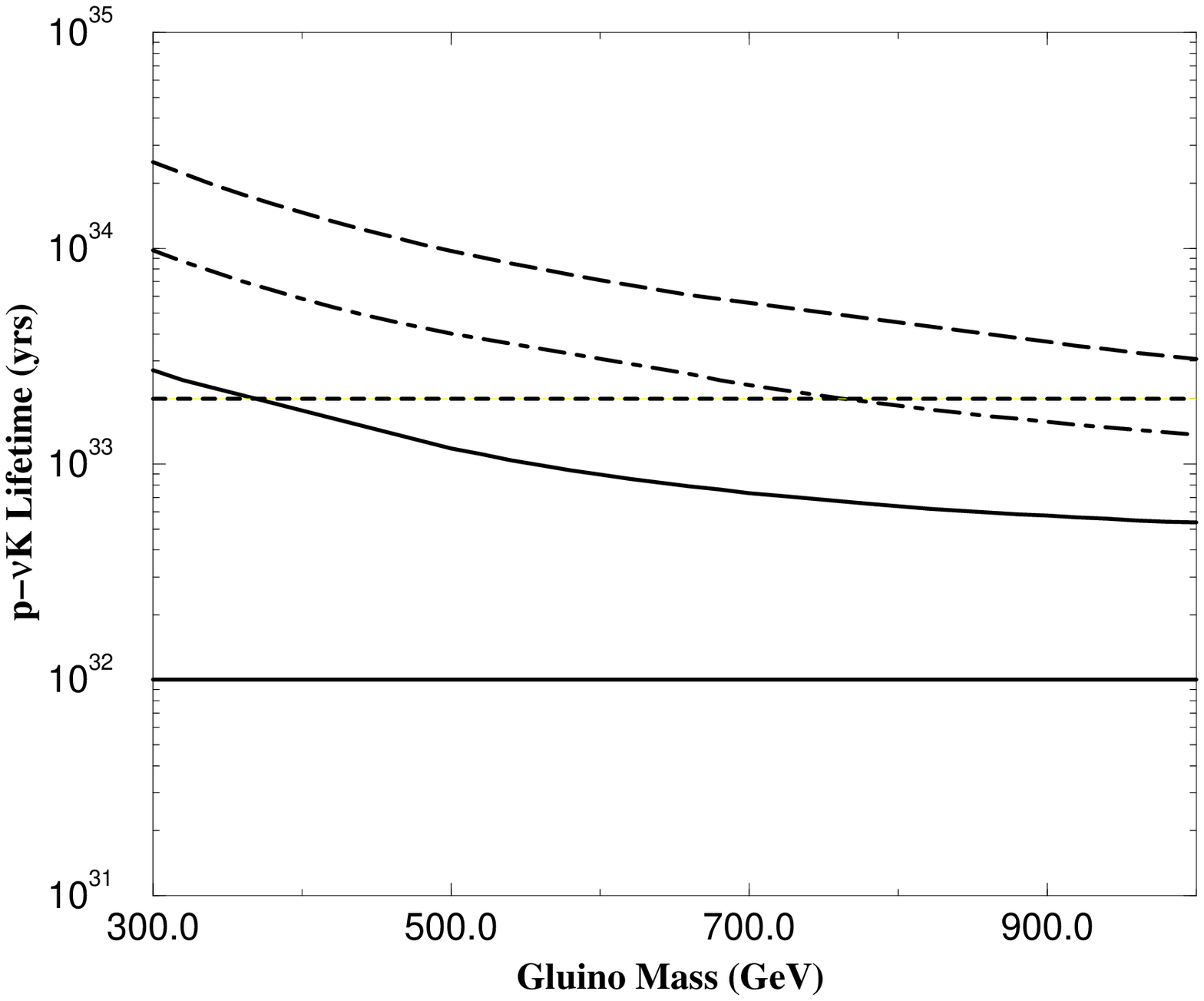}         %fig5 10/3/97  become fig6 10/6
	\small{ 
	Fig.1. The maximum $\tau(p\rightarrow \bar \nu K)$ lifetime
	in minimal SU(5) 
	supergravity unification for universal soft breaking as a function
	of the gluino mass for naturlaness limits on $m_0$ of 
	1 TeV (solid), 1.5 TeV (dashed-dot) and 2 TeV (dashed). The 
	solid horizontal line is the current experimental lower limit for
	this mode and the horizontal dashed line is the lower limit 
	expected from Super K.    	}
\end{minipage}
\end{figure}
%%%%%%%%%%%%%%%%%%%%%%%%%%%%%%%%%%%%%%%%%%%%%%%%%%%%%%%%%%%%%%%%%%%%%  

%%%%%%%%%%%%%%%%%%%%%%%%%%%%%%%%%%%%%%%%%%%%%%%%%%%%%%%%%%%%%%%%%%%%%  
%\begin{figure}[htbp]
%\begin{minipage}[t]{6.0in}
%	\includegraphics[angle=0,width=6.0in]{fig2.ps}         %fig5 10/3/97  become fig6 10/6
%	\small{ 
%	Fig.2. The maximum $\tau(p\rightarrow \bar \nu K)$ lifetime
%	in supergravity unification for the universal case as a function
%	of the gluino mass for naturalness limits on $m_0$ of 
%	1 TeV without relic density constraint (solid), 
%	 with  the  constraint $\Omega h^2<1$ (dashed) and with the
%	 constraint   $0.1<\Omega h^2<0.4$ (dashed-dot).  The 
%	 horizontal lines are as in Fig.1.  
% 	}
%\end{minipage}
%\end{figure}
%%%%%%%%%%%%%%%%%%%%%%%%%%%%%%%%%%%%%%%%%%%%%%%%%%%%%%%%%%%%%%%%%%%%%%  

%%%%%%%%%%%%%%%%%%%%%%%%%%%%%%%%%%%%%%%%%%%%%%%%%%%%%%%%%%%%%%%%%%%%%  
\begin{figure}[htbp]
\begin{minipage}[t]{6.0in}
	\includegraphics[angle=0,width=6.0in]{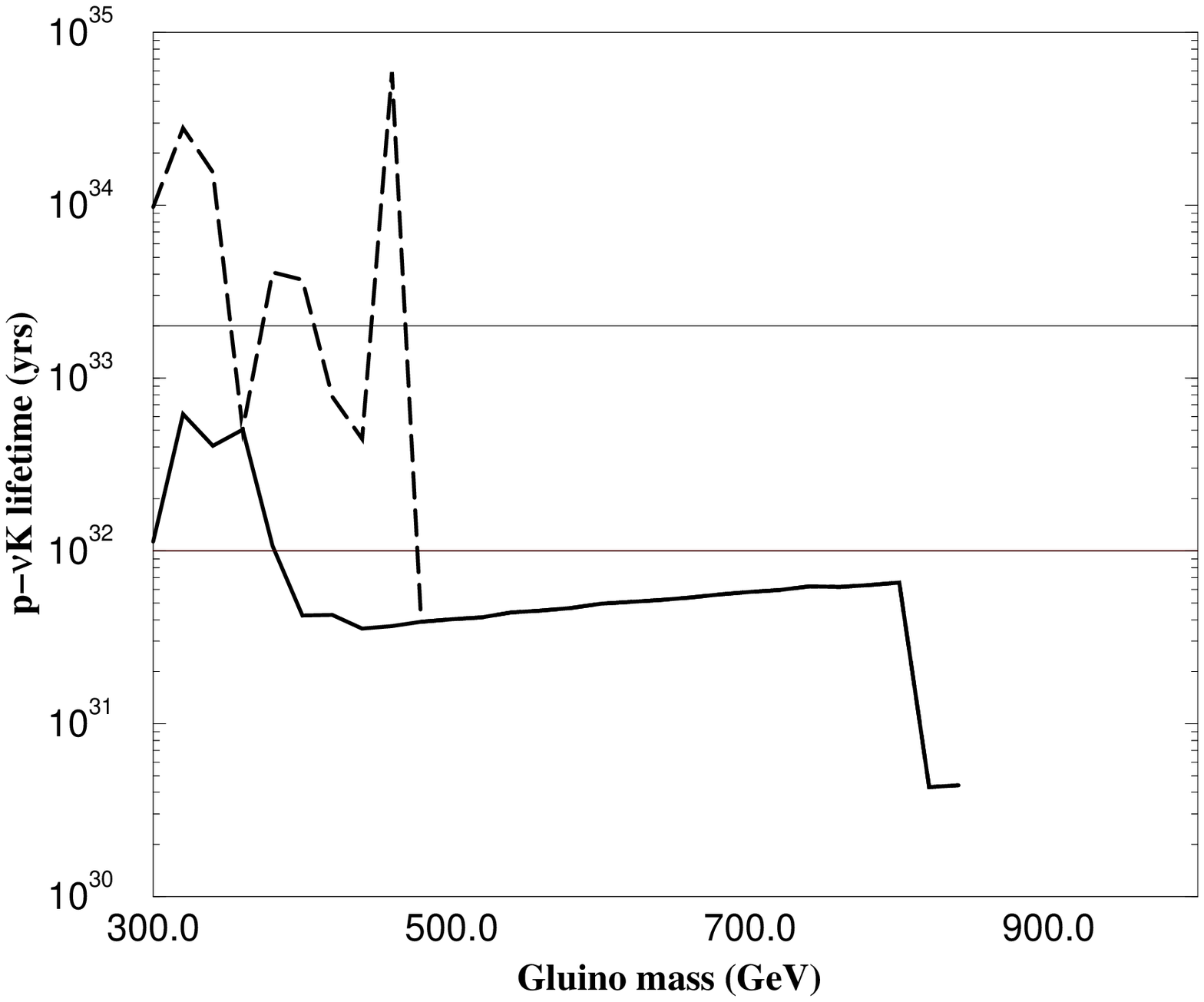}         %fig5 10/3/97  become fig6 10/6
	\small{ 
	Fig.2. The maximum $\tau(p\rightarrow \bar \nu K)$ lifetime
	in minimal SU(5) 
	supergravity unification for universal soft breaking as a function
	of the gluino mass  with the constraint $0.1<\Omega h^2<0.4$ 
	 for naturalness limits on $m_0$ of 1 TeV (solid),
%	 1.5 TeV (dashed-dot), 2 TeV (dotted)
	 and 5 TeV(dashed).  The 
	 horizontal lines are as in Fig.1.  
 	}
	
\end{minipage}
\end{figure}
%%%%%%%%%%%%%%%%%%%%%%%%%%%%%%%%%%%%%%%%%%%%%%%%%%%%%%%%%%%%%%%%%%%%%  

\begin{figure}[htbp]
\begin{minipage}[t]{6.0in}
	\includegraphics[angle=0,width=6.0in]{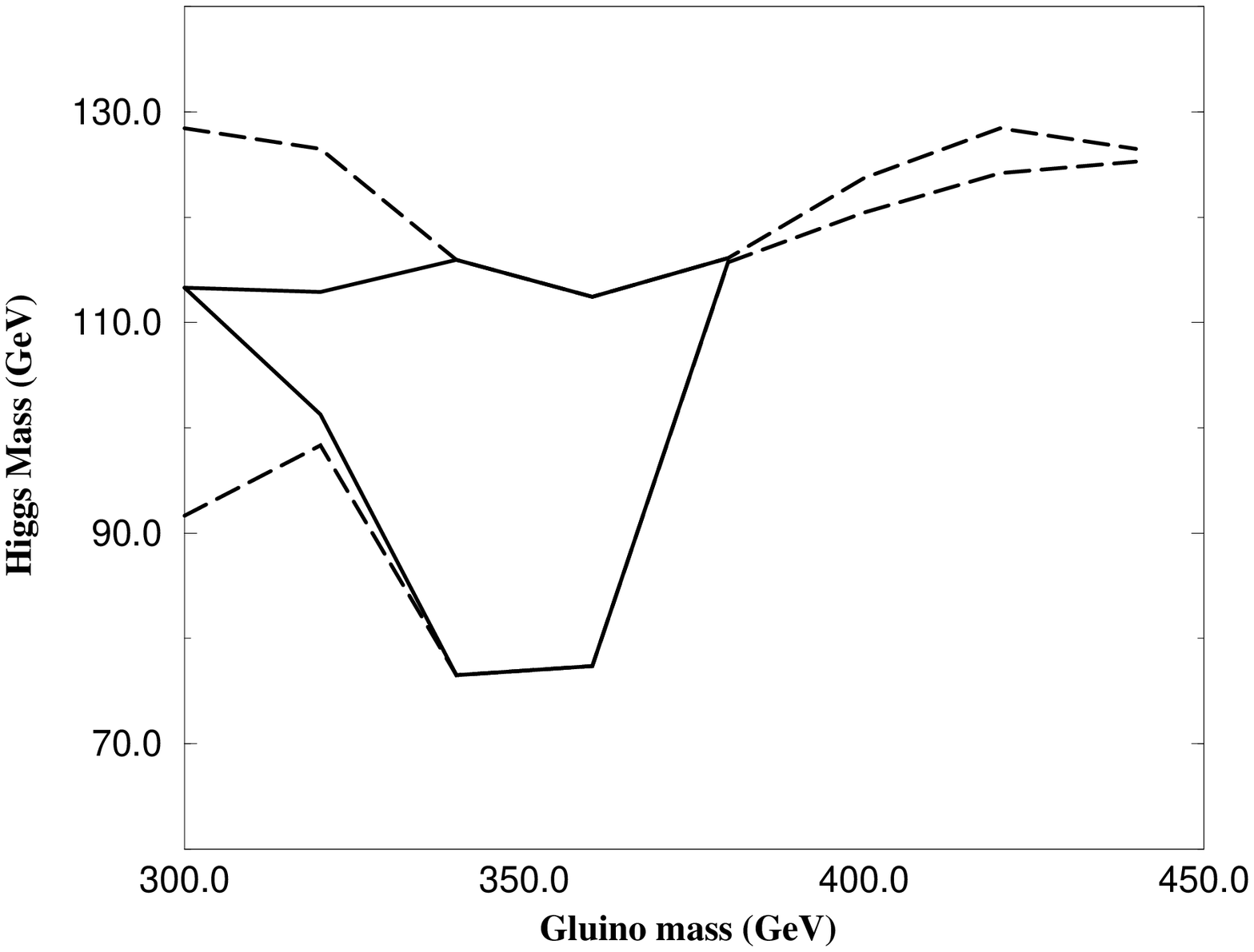}        
	\small{ 
	Fig.3. The maximum and the minimum of the Higgs mass as 
	a function of the gluino mass when the p-decay lifetime 
	constraint $\tau(p\rightarrow \bar \nu K)>1\times 10^{32}$ yr
	and the relic density constraint $0.1<\Omega h^2<0.4$ are 
	imposed in minimal supergravity with naturalness constraints 
	on $m_0$ of 1 TeV (solid) and 2 TeV(dashed).
 	}

\end{minipage}
\end{figure}

%%%%%%%%%%%%%%%%%%%%%%%%%%%%%%%%%%%%%%%%%%
%%%%%%%%%%%%%%%%%%%%%%%%%%%%%%%%%%%%%%%%%%%%%%%%%%%%%%%%%%%%%%%%%%%%%  
%\begin{figure}[htbp]
%\begin{minipage}[t]{6.0in}
%	\includegraphics[angle=0,width=6.0in]{fig3.eps}         
%	\small{ 
%	Fig.3. The maximum and the minimum of the Higgs mass as 
%	a function of the gluino mass when the p-decay lifetime 
%	constraint $\tau(p\rightarrow \bar \nu K)>1\times 10^{32}$ yr
%	and the relic den sity constraint $0.1<\Omega h^2<0.4$ are 
%	imposed in minimal supergravity with naturalness constraints 
%	on $m_0$ of 1 TeV (solid) and 2 TeV(dashed).
% 	}
%	
%\end{minipage}
%\end{figure}
%%%%%%%%%%%%%%%%%%%%%%%%%%%%%%%%%%%%%%%%%%%%%%%%%%%%%%%%%%%%%%%%%%%%%  
%%%%%%%%%%%%%%%%%%%%%%%%%%%%%%%%%%%%%%%%%%%%%%%%%%%%%%%%%%%%%%%%

%%%%%%%%%%%%%%%%%%%%%%%%%%%%%%%%%%%%%%%%%%%%%%%%%%%%%%%%%%%%%%%%%%%%%  
\begin{figure}[htbp]
\begin{minipage}[t]{6.0in}
	\includegraphics[angle=0,width=6.0in]{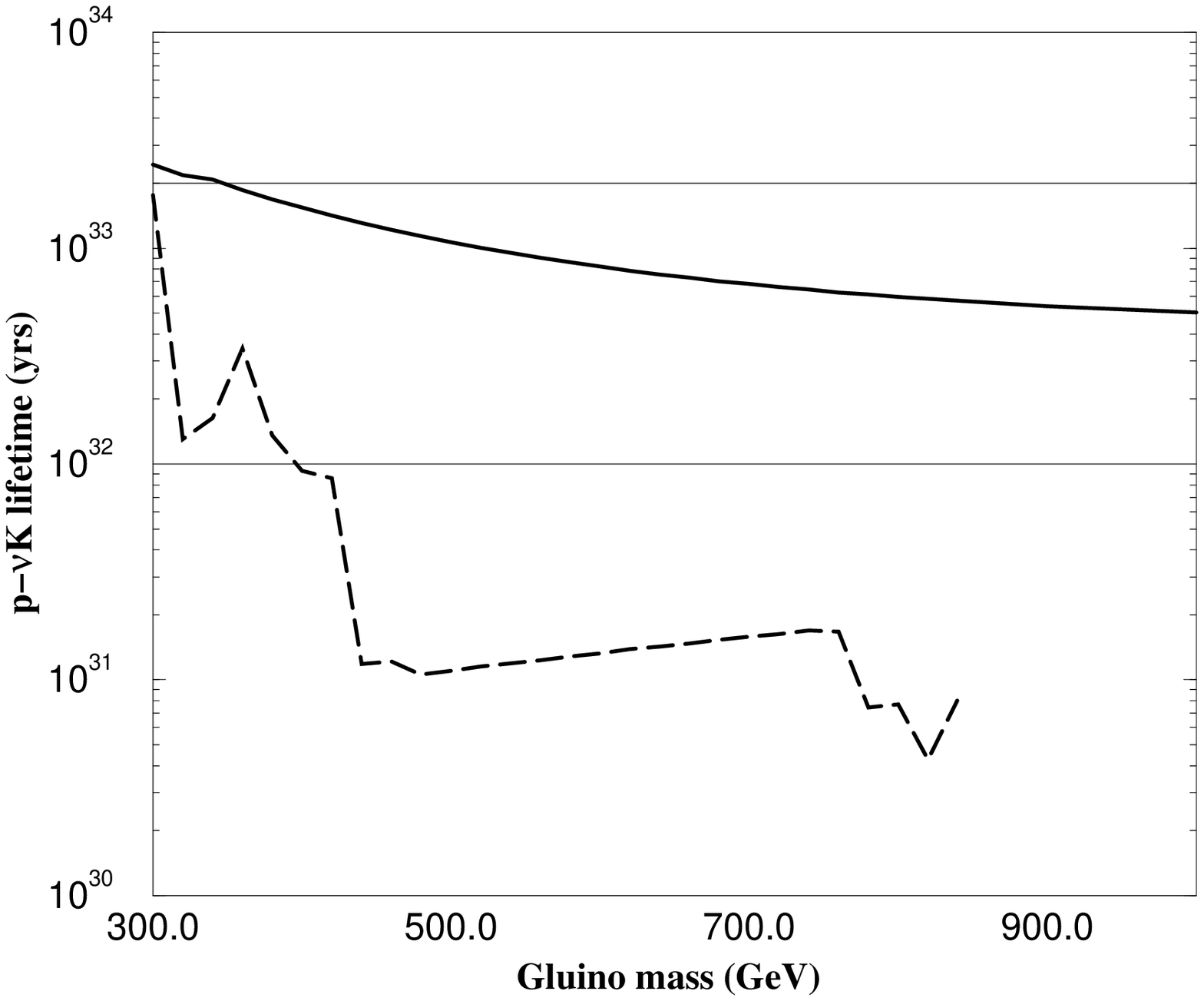}         %fig5 10/3/97  become fig6 10/6
	\small{ 
	Fig.4.  The maximum $\tau(p\rightarrow \bar \nu K)$ lifetime
	in minimal SU(5) supergravity unification for the non-universal case 
	$\delta_1=1=-\delta_2$ with limit on $m_0$ of $m_0\leq 1$ TeV
	 as a function of the gluino mass with no constraint on
	 the relic density (solid), 
%	 and with $\Omega h^2<1$ (dashed-dot),
	 and with   $0.1\leq \Omega h^2<0.4$ (dashed). The 
	 horizontal lines are as in Fig.1.  
 	}
	
\end{minipage}
\end{figure}
%%%%%%%%%%%%%%%%%%%%%%%%%%%%%%%%%%%%%%%%%%%%%%%%%%%%%%%%%%%%%%%%%%%%%  
%------------------------------------------------------------------

%%%%%%%%%%%%%%%%%%%%%%%%%%%%%%%%%%%%%%%%%%%%%%%%%%%%%%%%%%%%%%%%%%%%%  
\begin{figure}[htbp]
\begin{minipage}[t]{6.0in}
	\includegraphics[angle=0,width=6.0in]{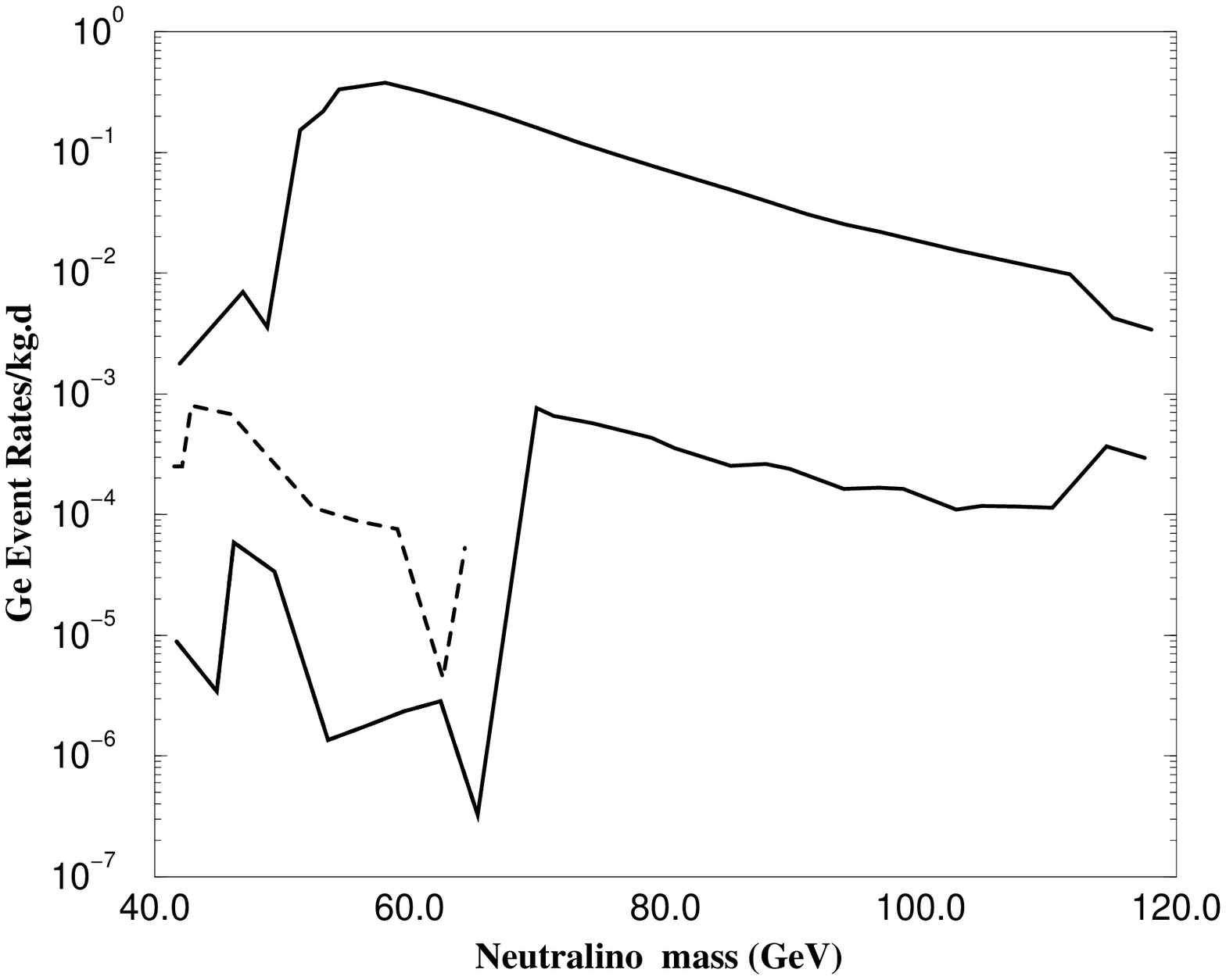}  
	\small{ 
	Fig.5. The maximum and minimum of event rates for the scattering 
	of  neutralinos off germanium target as a function of the 
	neutralino mass with 
	the relic density constraint $0.1<\Omega h^2<0.4$ when the
	naturalness constraint on $m_0$ is 1 TeV for the cases 
	(1) without p decay 
	constraint (solid), and (2)
	under the proton lifetime constraint 
	$\tau (p\rightarrow \bar\nu K)>1\times 10^{32}$ yr (dashed)
	(The  lower dashed curve coincides with the lower solid curve
	and is thus not visible).
 	}
	
\end{minipage}
\end{figure}


\newpage
\begin{thebibliography}{9}

\bibitem{wein} S. Weinberg,~Phys. Rev. {\bf D26} (1982) 287; 
N. Sakai and T. Yanagida, Nucl. Phys.{\bf B197} (1982) 533; 
S. Dimopoulos, S. Raby  and F. Wilczek, Phys.Lett.
 {\bf 112B} (1982) 133;
J. Ellis, D.V. Nanopoulos and S. Rudaz, Nucl. Phys.
{\bf  B202} (1982) 43;
B.A. Campbell, J. Ellis and D.V. Nanopoulos,
 Phys. Lett. {\bf 141B}  (1984) 299;
S. Chadha, G.D. Coughlan, M. Daniel
 and G.G. Ross, Phys. Lett.{\bf 149B} (1984) 47;
 R. Arnowitt, A.H. Chamseddine and P. Nath, Phys. Lett.
{\bf 156B} (1985) 215;
P. Nath, R. Arnowitt and A.H. Chamseddine, Phys. Rev. {\bf 32D}
 (1985) 2348;
 J. Hisano, H. Murayama and T. Yanagida, Nucl. Phys.
{\bf B402} (1993) 46.
\bibitem{acn}
 R. Arnowitt and P. Nath, Phys. Rev. {\bf 49}
(1994) 1479.

\bibitem{can}
A.H. Chamseddine, R. Arnowitt and P. Nath, Phys. Rev. Lett {\bf 29}
(1982) 970.
\bibitem{applied}
For reviews see P.Nath,   R. Arnowitt and A.H.Chamseddine ,
``Applied N=1 Supergravity" (World Scientific,
Singapore, 1984);
 H.P. Nilles, Phys. Rep. {\bf 110} (1984) 1; R. Arnowitt and
P. Nath, Proc of VII J.A. Swieca Summer School (World Scientific, Singapore
1994).

\bibitem{scaling}
R. Arnowitt and P. Nath, Phys. Rev. Lett. {\bf 69} (1992) 725;
P. Nath and R.
Arnowitt, Phys. Lett. {\bf B289} (1992) 368.



\bibitem{soni}
S. K. Soni and H. A. Weldon, Phys. Lett. {\bf B126} (1983) 215; 
V. S. Kaplunovsky
and J. Louis, Phys. Lett. {\bf B306} (1993) 268.



\bibitem{matallio}
D. Matalliotakis and H.P. Nilles, Nucl.Phys.{\bf B435},
(1995) 115; M. Olechowski and S. Pokorski, Phys.Lett. {\bf B344}, 
(1995) 201;
N. Polonski and A. Pomarol,Phys.Rev.{\bf D51} (1995) 6532;
V. Berezinsky, A. Bottino, J. Ellis, N. Forrengo,
G. Mignola, and S. Scopel, Astropart. Phys.{\bf 5} (1996) 1;ibid,
{\bf  5} (1996) 333.

\bibitem{nonuni}
P. Nath and R. Arnowitt, Phys. Rev. {\bf D56} (1997) 2820;
 R.~Arnowitt and P.~Nath, Phys.
Rev.{\bf D56} (1997) 2833.

\bibitem{flipped}
An exception is the flipped SU(5) model, I. Antoniadis, J. Ellis, J.S.
Hagelin and D.V. Nanopoulos, Phys. Lett. {\bf B231} (1987) 65. 

\bibitem{gavela} M.B.Gavela et al, Nucl.Phys.{\bf B312} (1989) 269.


\bibitem{pdg} Particle Data Group, Phys.Rev. {\bf D54} (1996) 1.



\bibitem{totsuka} Y.Totsuka, Proc. XXIV Conf. on High Energy Physics,
Munich, 1988,Eds. R.Kotthaus and J.H. Kuhn (Springer Verlag, Berlin, 
Heidelberg,1989).


\bibitem{icarus} ICARUS Detector Group, Int. Symposium on Neutrino 
Astrophsyics, Takayama. 1992.


\bibitem{jungman}
For a review see G. Jungman, M. Kamionkowski and K. 
Greist,  Phys. Rep. {\bf 267} (1995) 195;
   E.W. Kolb and M.S. Turner, ``The
Early Universe'' (Addison-Wesley, Redwood City, 1989);
P. Nath and R. Arnowitt, Proc. of the Workshop on Aspects of Dark Matter
in Astrophysics and Particle Physics, Heidelberg, Germany 16-20 
September, 1996, World Scientific p. 333.


\bibitem{greist}
K. Greist and D. Seckel, Phys. Rev. {\bf D43} (1991) 3191; P. Gondolo and G.
Gelmini, Nucl. Phys. {\bf B360} (1991) 145.


\bibitem{accurate}
R. Arnowitt and P. Nath, Phys. Lett. {\bf B299} (1993) 103; 
 Phys. Rev. Lett. {\bf 70} (1993) 3696; Phys. Rev. {\bf D54} (1996) 2374;
M. Drees and A. Yamada, Phys. Rev. {\bf D53} (1996) 1586;
 H. Baer and M. Brhlick, Phys. Rev. {\bf D53} (1996) 597; 
 V. Barger and C. Kao, hep-ph/9704403.


\bibitem{noscale}
P. Nath and R. Arnowitt, Phys. Lett. {\bf B289} (1992) 308. The analysis
of this work of course  does not apply to the flipped no-scale 
models\cite{flipped} which suppress p decay by a change of the choice of matter 
representations.


\bibitem{barr}
K.S. Babu and S.M. Barr, Phys. Rev. {\bf D51}(1995)2463.

\bibitem{lucas}
V. Lucas and S. Raby, Phys. Rev. {\bf D54} (1996) 2261; ibid. {\bf D55}
(1997) 6986.


\bibitem{urano}
S. Urano and R. Arnowitt, hep-ph/9611389.


\bibitem{georgi}
H. Georgi and C. Jarlskog, Phys. Lett. {\bf B86} (1979) 297;
J. Harvey, P. Ramond and D. Reiss, Phys. Lett. {\bf B92}
(1980) 309; P. Ramond, R.G. Roberts,
G.G. Ross, Nucl. Phys. {\bf B406} (1993) 19.

\bibitem{anderson}
 G. Anderson, S. Raby, S. Dimopoulos, L. Hall, and G.D.
Starkman, Phys. Rev. {\bf D49} (1994) 3660.

\bibitem{pn}
P. Nath, 
  Phys. Rev. Lett. {\bf 76}  (1996) 2218. 


\bibitem{calabi}
R. Arnowitt and P. Nath. Phys. Rev. Lett. {\bf 62} (1989) 2225.

\bibitem{cline}
For a discussion on the possibilites of more sensitive detectors see
e.g., D. Cline, Nucl. Phys. B(Proc. Suppl.) {\bf 51B} (1996) 304, and
H. Klapdor-Kleingrothaus, in Proc. of "Non-accelerator New Physics",
Dubna, July 7-11, 1997. 



\end{thebibliography}
\end{document}